\documentclass{sf2a-conf2024}
\usepackage{graphicx}
\usepackage{hyperref}
\usepackage[]{natbib}  
\usepackage{epstopdf}

\def\BibTeX{{\rm B\kern-.05em{\sc i\kern-.025em b}\kern-.08em
    T\kern-.1667em\lower.7ex\hbox{E}\kern-.125emX}}
\bibpunct{(}{)}{;}{a}{}{,}  

\begin{document}

\TitreGlobal{SF2A 2024}


\title{Mass \& Light in Galaxy Clusters:\\ Parametric strong-lensing approach}

\runningtitle{Mass \& Light in Galaxy Clusters}

\author{Marceau Limousin}\address{Aix Marseille Univ, CNRS, CNES, LAM, Marseille, France.}

\setcounter{page}{237}


\maketitle


\begin{abstract}
In the cold dark matter paradigm, the association between the hypothetic dark matter and its stellar counterpart is
expected.
However, parametric strong lensing studies of galaxy clusters often display "misleading features":
group/cluster scale dark matter components
without any stellar counterpart, offsets between both components larger
than what might be allowed by neither Cold Dark Matter nor self interacting 
Dark Matter models,
or significant unexplained external shear components.
I am revisiting mass models where such "misleading" (and interesting) features have
been reported, adopting the following working hypothesis: any group or cluster scale
dark matter clump introduced in the modelling should be associated with a luminous
counterpart, and any well motivated and reliable prior
should be considered, even when this degrades the fit.
The goal is to derive a physically motivated description of the dark matter
component which might be compared to theoretical expectations.
I succeed doing so in galaxy clusters AS\,1063, MACS\,J0416 and MACS\,J1206, 
finding that the shape of the inner dark matter component has a flat density profile.
These findings may be useful for the interpretation within dark matter scenario,
such as self-interacting dark matter.
I fail in Abell\,370: a three dark matter clumps mass model (each clump being
associated with its stellar counterpart) is unable to reproduce the observational 
constraints with a precision smaller than 2.3".
In order to provide a sub-arcsec precision, I need to describe the dark matter
distribution using a four dark matter clumps model, one having no 
stellar counterpart, and another one presenting a significant offset with its 
associated stellar counterpart, as found in earlier works.
Investigating this solution, I present a class of such models which can accurately
reproduce the multiple images, but whose parameters for the dark matter component
are poorly constrained, limiting any insights on its properties.
Examining the \emph{total} projected mass maps, I however find a good agreement 
between the total mass and the stellar distribution in Abell\,370, both being, to 
first order, bimodal. 
I interpret the "misleading features" of the four dark matter clumps mass model and the failure of the three
dark matter clumps mass model as being symptomatic of the lack of realism of a parametric description
of the dark matter distribution in such a complex merging cluster.
I encourage caution and criticism on the
outputs of parametric strong lensing modelling.

\end{abstract}



\section{Introduction}

Dark matter (DM) is an elusive component that is thought to largely dominate the mass budget
in astrophysical objects over a wide range of scales, in particular in galaxy clusters.
However, more than 80 years after the first indirect evidence for DM in galaxy clusters (Zwicky 1937), we have no definitive
clues about its existence, even though it is sometimes taken for granted.
Evidence for DM is indirect only, and no well-understood and characterised particle detector has
detected it so far, despite intense effort of the community.
As long as no such direct detection is reliably achieved, DM remains, from my point
of view, a hypothesis.

Both observations and numerical simulations do support the association
between DM and light (the associated stellar component, in most cases in the 
form of the brightest cluster galaxy (BCG) at the galaxy cluster scale).
Observationally, no cluster scale DM clump without any associated light
concentration has been reliably detected so far.
Besides, in hydrodynamical simulations, stars do form in the potentiel well
of DM halos. This results into the hierarchical formation of a bright galaxy
found at the centre of the underlying DM halo.

If DM is collisionless as proposed in the Cold Dark Matter (CDM) scenario,
the association between mass and light should be perfect,
\emph{i.e.} the offset between the peaks of each component should be
equal to 0 \citep{Roche_2024}.
If DM is self interacting (SIDM), such an offset is possible and should be at most
of the order of a few dozen of kpc, according to simulations
\citep[\emph{e.g.}][]{Kim_2017,SIDM_review,Adhikari_2022}.

Strong lensing (SL) is an essential probe of the DM distribution in the centre of
galaxy clusters, where the mass density is so high that space time is locally deformed such that
multiple images of background sources can form. This provides valuable constraints on the \emph{projected} mass distribution.

Parametric SL mass modelling relies on the following working hypothesis, supported by N-body simulations:
A galaxy cluster is an object composed of different mass clumps. 
Each component is associated
with a luminous counterpart and can (to some extent) be described
parametrically.
One advantage of parametric SL modelling is that the description of these mass clumps
can be directly compared with theoretical expectations.
However, a parametric description is sometimes not accurate nor adapted, emphasising the
limit of parametric mass modelling and the need for more flexible approaches.
We usually consider two types of mass clumps
in parametric SL modelling: cluster-scale DM clumps (whose typical projected mass within
a 50" aperture is about 10$^{14}$M$_{\mathrm{O}}$ at $z\sim$\,0.2) and galaxy-scale DM clumps associated with
individual galaxies. Added to this description of the dominant DM component, the
mass component associated with the X-ray gas can also be considered \citep{Bonamigo_2018,Beauchesne24}.

Parametric SL modelling displays interesting and puzzling features that can be
"misleading". These are the following:

\begin{itemize}
\item First, it sometimes requires DM clumps whose position does not coincide with that of any
luminous counterpart.
This is the case in complicated merging clusters, like Abell~370, but also in
apparently unimodal clusters as MACS\,J1206.
These dark clumps are usually added in order to improve the fit
significantly, but the physical interpretation of these clumps is not straightforward. Moreover, when they are taken for
granted, their inclusion in the mass budget might be misleading. We might wonder whether we are really
witnessing a dark clump.

\item Second, offsets between DM clumps and the associated light peak are reported,
larger than what might be allowed by SIDM scenarios.

\item Third, some mass models require a non-negligible external shear component 
($\gamma_{\rm ext}$) in order to significantly
improve the goodness of the reconstruction, but the physical origin of this external shear is not always clear.

\item Finally, large core radii (sometimes larger than 100\,kpc) are sometimes reported in parametric SL studies
\citep[e.g.][]{Newman_2013,Lagattuta_2019,Richard_2021}. 
A thorough SIDM investigation of the size of the core radii in the galaxy cluster
mass regime is still lacking,
but we have some indications about its order of magnitude,
which should be smaller than $\sim$50\,kpc 
\citep{Rocha_2013,Robertson_2017,Fischer_2021}.

\end{itemize}

From these considerations, I have been revisiting some mass models where former
studies reported "misleading features", within the following working assumption:
any group or cluster scale DM clump introduced in the modelling should be
associated with a luminous counterpart (within the SIDM allowance), and any
well motivated and reliable, observationnaly motivated prior should be considered, 
even when this degrades the fit, quantified by the root mean square 
between the observed and model-generated images.
The goal is to see if I can get rid of these features, present a physically
motivated model and probe DM physics, in particular its inner slope.
For each cluster I have studied, I list the misleading features present in former
works and I summarize the main results, which can be found in the
following papers: \citet{Limousin22} and \citet{Limousin_2024}.
A former paper \citep{Limousin_2016} might be of interest when it comes to degeneracies
in parametric SL modelling.

\section{AS\,1063}
AS\,1063 is the only Hubble Frontier Fields \citep[HFF,][]{LotzHFF} cluster that appears
to be unimodal and dominated by a BCG, with which a DM halo is associated.
However, a galaxy group generating additional SL features is located in the 
north-east.
Former study by \citet[][B19 hereafter]{Bergamini_2019} associate a mass clump with this
group, but its location does not coincide at all with it (separation of 40").
I did revisit the B19 mass model, forcing the position of this second mass clump to
coincide with the light distribution of the most luminous galaxy of this group.
I obtained an RMS of 0.67" (versus 0.55" reported by B19).
The main DM component has a core radius equal to 89$\pm$5\,kpc.
I then forced the core radius to be smaller than 10\,kpc and redid the modelling, 
in order to see if a non cored mass model can reproduce the SL constraints.
The RMS goes up to 3.83".
This suggest that a cored DM profile is favoured in AS\,1063.

\section{MACS\,J0416}

The HFF cluster MACS\,J0416 is multimodal, with three well defined
light peaks.
The last study by \citet{Bergamini_2021} reproduced 182 multiple images using a 4 DM mass
clumps, reaching an RMS of 0.40".
Two of these DM mass clumps are associated with a light peak, and two are
located in the south-west of the cluster core, none of which being clearly
associated with the brightest galaxy located in this area.
Considering a three DM mass clumps model, each being associated with a light peak,
I reached an RMS of 0.63". The core radii of each clump are larger than 50\,kpc.
If I impose them to be smaller than 10\,kpc, the RMS goes to 2.07".
Therefore, a cored mass model is preferred.

\section{MACS\,J1206}

MACS\,J1206 is a unimodal cD dominated galaxy cluster.
Former works (\emph{e.g.} B19) reproduced 82 multiple images using a mass
distribution composed of three DM haloes and a strong external shear 
($\gamma_{\rm ext}$=0.12), reporting an RMS equal to 0.46".
If one of these mass clumps is coincident with the cD galaxy, the others are not
associated with any luminous counterpart, and are claimed to be necessary to
reproduce the apparently elongated asymmetry of the cluster.
Using a single DM clump associated with the cD galaxy does not allow to reach a
decent fit (RMS=2.24"), suggesting that MACS\,J1206 cannot be reliably
described by a pure parametric model in which each mass component would be
associated with a luminous counterpart.

In a second step, I added a (mild) perturbation to the parametric modelling to
see whether this might help to provide a decent fit. This perturbation consists
of a surface of 2D B-spline functions that are added to the lensing
potential \citep{Beauchesne_2021}.
First, I verified that this perturbation is mild enough to avoid modifying
the parameters of the associated parametric mass model significantly, which
would make us lose the advantages of the parametric mass modelling.
I tested the inclusion of these perturbations on AS\,1063, which is already well 
described by a parametric mass model, looking at the response of the reference
mass model to this perturbation.
The RMS is improved, and the parameters of the reference model do agree within
the 3$\sigma$ error bars with the parameters obtained with the perturbation.

Encouraged by this test, I reconsidered the mass model of MACS\,J1206 using
a single DM mass clump and the perturbation, reaching an RMS of 0.53".
Moreover, no external shear is required.
The core radius of this DM mass clump equals to 57\,kpc.
If I impose it to be smaller than 10\,kpc, the resulting RMS is 7",
which definitely favors a cored mass model.
 
\section{Abell~370}

Abell\,370 is a multimodal merging HFF cluster. 
The light distribution is dominated by the light associated with two
dominant bright galaxies (BCG-N and BCG-S). We also observe a light 
concentration in the east/north-east.

Former SL studies described Abell~370 
parametrically by a four dark matter clumps model, 
as well as a significant external shear component, which physical origin 
remained a challenge. The dark matter distribution features a mass clump with 
no stellar counterpart located between BCG-N and BCG-S, and a significant 
offset (larger than what is allowed by SIDM) between the northern dark matter clump and 
its associated stellar counterpart.
I began by revisiting this mass model. Sampling this complex parameter space 
with \textsc{mcmc} techniques, I found a four dark matter clumps solution which 
does not require any external shear and provides a slightly better RMS compared 
to previous models (0.7" compared to 0.9"). Investigating further this new 
solution, in particular playing with the parameters leading the \textsc{mcmc} 
sampler, I presented a class of models which can accurately reproduce the 
strong lensing data, but whose parameters for the dark matter component 
are poorly constrained, limiting any insights on its properties, in particular
its inner shape.

I then investigated a model where each large scale dark matter clump 
is associated with
a stellar counterpart. This three dark matter clumps model is unable to
reproduce the observational constraints with an RMS smaller than 2.3",
and the parameters describing this dark matter component are also 
poorly constrained.
The addition of a B-spline perturbation did not help.

Still, the total projected mass is well constrained and is linked to the 
stellar component, the two main \emph{total mass peaks}
being coincident with the two BCGs.
I therefore concluded that the \emph{total mass is traced by light in 
Abell~370.}
Having said that, it is relevant to discuss what is learned about 
the underlying DM distribution, which, taken as such, might be misleading. What is the interpretation of this ”dark clump”: are we detecting a
”dark clump”? What is the interpretation of the offset between the 
Northern DM clump and BCG-N, which is larger than what might be allowed by SIDM?
These interesting features are clearly required by the data in order 
to reproduce the observed positions of the multiple images with a 
sub-arcsecond precision. I do interpret these as not being ”real” 
but rather being necessary to compensate the lack of reality of the 
parametric description of DM clumps during a cluster merging process.
Indeed, the DM component is described using idealised parametric mass 
profiles (e.g. dPIE or NFW). This description, though simple,
can be reliable, which is remarkable. This is the case in AS\,1063, MACS\,J1206 and MACS\,J0416.
In Abell\,370, such a simple description of the different DM components 
involved in the merging process might not fully capture the complex 
underlying physics, hence, some features, as the ones reported here, 
are needed to account for the deviations from our idealised 
parametric descriptions.

\section{Conclusions}
Overall, this analysis suggest evidence for cored cluster-scale dark matter haloes in
the three clusters for which I have been able to propose a model where each DM clump is
associated with a luminous counterpart.
These findings may be useful for the interpretation within alternative dark matter scenario,
such as self-interacting dark matter.

Even in the JWST era, where hundreds of multiple images are observed, SL mass 
reconstructions still suffer from degeneracies, in particular in merging clusters, and 
caution and criticism should be taken when reading and interpreting the results of any SL model. 
Furthermore, authors could discuss more the limitations of their models, and help the reader 
to understand and interpret their results. We therefore encourage caution and criticism on the 
outputs of parametric SL modelling.

These results also have some implications for high redshift studies using clusters as
natural telescopes.

\begin{acknowledgements}
I thank the SOC and the LOC (!) for organizing such a lovely and inspiring meeting.
I acknowledge CNRS and CNES for support.
The mass models discussed here have been performed using facilities offered by CeSAM (Centre de 
donn\'eeS Astrophysique de Marseille).
\end{acknowledgements}

\bibliographystyle{aa}  
\bibliography{Limousin_S17} 

\end{document}